\documentstyle[12pt,aaspp]{article}

\def\jtwoone{{\sl J} = $2 {\rightarrow} 1$}
\def\jonezero{{\sl J} = $1 {\rightarrow} 0$}
\def\etal{{\rm et al.}}

\def\rosat{{\it ROSAT}}
\def\einstein{{\it Einstein}}
\def\myarcmin{^\prime\mskip-5mu}
\def\hii{H {\sc{ii}}}

\begin{document}

\received{13 August 1996}
\accepted{9 December 1996}

\slugcomment{Scheduled to appear in {\it Ap.~J.}, Vol.~480,  May 10, 1997}

\title{Supernova Remnants Associated with Molecular Clouds in
the Large Magellanic Cloud}

\author{
   Kenneth R. Banas\altaffilmark{1,2}, 
   John P.~Hughes\altaffilmark{1,3},
   L.~Bronfman\altaffilmark{4}, and
   L.-\AA.~Nyman\altaffilmark{5,6}}
\altaffiltext{1}{Harvard-Smithsonian Center for Astrophysics, 60 Garden Street,
Cambridge, MA 02138}
\altaffiltext{2}{krb@astro.caltech.edu, California Institute of
Technology, Department of Astronomy, 105-24, Pasadena, CA 91125}
\altaffiltext{3}{jph@physics.rutgers.edu, Department of Physics and
Astronomy, Rutgers University, P.O.~Box 849, Piscataway, NJ 08855-0849}
\altaffiltext{4}{Departamento de Astronom\'ia, Universidad de Chile,
Casilla 36-D, Santiago, Chile}
\altaffiltext{5}{Onsala Space Observatory, S-439 92 Onsala, Sweden}
\altaffiltext{6}{SEST, ESO, Casilla 19001, Santiago 19, Chile}


\begin{abstract}

We used the Swedish-ESO Submillimeter Telescope (SEST) to search for
CO emission associated with three supernova remnants (SNRs) in the
Large Magellanic Cloud: N49, N132D, and N23. Observations were carried
out in the \jtwoone\ rotational transition of CO (230.5 GHz) where the
half power beamwidth of the SEST is 23$^{\prime\prime}$. Molecular
clouds were discovered near N49 and N132D; no CO emission was
discovered in the region we mapped near N23. The N49 cloud has a peak
line temperature of 0.75 K, spatial scale of $\sim$7 pc and virial
mass of $\sim$$3\times 10^4\, M_\odot$. The N132D cloud is brighter with a
peak temperature of 5 K; it is also larger $\sim$22 pc and
considerably more massive $\sim$$2\times 10^5\, M_\odot$.  The velocities
derived for the clouds near N49 and N132D, $+$286.0 and $+$264.0 km
s$^{-1}$, agree well with the previously known velocities of the
associated SNRs: $+$286 km s$^{-1}$ and $+$268 km s$^{-1}$,
respectively. \rosat\ X-ray images show that the ambient density into
which the remnants are expanding appears to be significantly increased
in the direction of the clouds. Taken together these observations
indicate a physical association between the remnants and their
respective, presumably natal, molecular clouds.  The association of
N49 and N132D with dense regions of molecular material means that both
were likely products of short-lived progenitors that exploded as
core-collapse supernovae.

\end{abstract}

\section{Introduction}

Stellar evolutionary models predict massive stars will die in
supernova explosions near where they were born.  It follows that
supernova remnants (SNRs) associated with the molecular clouds from
which their progenitors were born must have arisen from massive star,
core-collapse supernovae. It is further believed that blast waves from
supernova acting on dense atomic and molecular gas can initiate star
formation as one part in a clearly cyclic process. Therefore the
identification and study of SNRs/molecular clouds associations is
important to our understanding of the structure and dynamics of the
interstellar medium (ISM), the ratio of thermonuclear to core-collapse
supernova, the life cycle of stars, the destruction of molecular
clouds, and so on.

The Large Magellanic Cloud (LMC) is well suited to the identification
and study of SNR and molecular cloud associations.  The Cloud is
nearby, relatively unobscured, and has been extensively observed at
nearly all wavebands.  A sizeable number ($>$30) of supernova remnants
have been identified in the LMC from X-ray, optical, and radio
measurements (Mathewson \etal\ 1983, 1984, 1985).  Cohen \etal\ (1988)
systematically surveyed the LMC for CO line emission and found a
general correspondence between the CO emission and such Population I
objects as SNRs and \hii\ regions. However, because of the limited
spatial resolution of the survey ($12^\prime$), a detailed association
between individual objects and molecular clouds was not
possible. Guided by the results of this survey, we selected three
probable SNR/molecular cloud associations for study at considerably
higher angular resolution using the Swedish-ESO Submillimeter
Telescope (SEST) at La Silla, Chile.  The angular resolution of the
SEST at 230 GHz, $\sim$20$^{\prime\prime}$, corresponds to a linear size
of 4.85 pc for a distance to the LMC of 50 kpc, which we adopt throughout.

The remnants we selected, in addition to being positionally coincident
with CO emission from the LMC survey, were required to show evidence
for enhanced X-ray and optical emission along part of the rim, which
might be indicative of a density gradient in the ambient medium. The
best candidates were the SNRs 0525$-$66.1, 0525$-$69.6, and
0506$-$68.0 (Mathewson \etal\ 1983), which we will refer to henceforth
using their Henize (1956) catalog names: N49, N132D, and N23,
respectively.

N49 shows the highest optical surface brightness of all LMC SNRs. Its
X-ray (and optical) morphology shows a sharp increase in emission
along the southeast edge of the remnant (Vancura \etal\ 1992a).  It
appears to be embedded within one of the brighter and more massive CO
clouds in the LMC (among clouds, that is, that are not in the 30
Doradus region).  N132D is the brightest soft X-ray emitting SNR in
the LMC (Mathewson \etal\ 1983) and optically it displays high
velocity oxygen-rich material.  Consequently it is one of the most
frequently studied of LMC remnants.  Toward the south, N132D has a
nearly circular, very bright limb, which is about three-quarters
complete, while the remaining quarter of the remnant appears as a
``blow-out'' toward the northeast.  Hughes (1987) interpreted this
structure as a result of the ionizing radiation and stellar wind of
the progenitor acting on a medium with a strong gradient in ambient
ISM density.  The LMC CO survey showed N132D sitting on the northern
edge of a relatively modest cloud.  N23 has the weakest X-ray emission
of the three chosen SNRs.  Its morphology is similar to that of N49
and N132D in that it shows an increase in emission along one limb, in
this case the eastern one.  The remnant sits on the southern boundary
of a modest CO cloud.

In the following, we describe our SEST observations, data reduction, and
error analysis (\S 2); give the results on the molecular clouds found
near N49 and N132D (\S 3); discuss these results in the context of
what else is known about these remnants (\S 4); estimate the masses of
the newly discovered molecular cloud (\S 5); and in \S6 we summarize.
A preliminary report on a subset of the SEST data was given in Hughes,
Bronfman, \& Nyman (1991).

\section{Observations and Analysis Techniques}
\subsection{Observations}

High resolution imaging of three candidate SNR/molecular cloud
associations was undertaken with the SEST on La Silla, Chile, on June
18-21, 1989 and again on October 16-19, 1990. The CO \jtwoone\ (at
230.5 GHz) observations were done with a linearly polarized Schottky
receiver giving a typical system temperature above the atmosphere of
800--1600 K, depending on elevation and weather conditions. At this
frequency the beamwidth of the telescope is 23\arcsec$\:$ (FWHM),
considerably better than the $12^\prime$ resolution of Cohen \etal\
(1988), thereby allowing us to resolve finer structure within the
molecular clouds near the SNRs. The main beam efficiency of the telescope,
$\eta_{mb}$, was 0.54 during the 1989 observations, and due to an
improvement in the surface accuracy in 1990, the beam efficiency
during the October 1990 observations was about 0.70. We quote results
in terms of the antenna temperature corrected for beam efficiency,
$T_{mb} = T_A^\star/\eta$. Spectra were taken with an acousto-optical
spectrometer with a bandwidth of 86 MHz, a channel separation of 43
kHz, and a resolution of 80 kHz, which at the observed frequency
corresponds to a velocity resolution of 0.1 km s$^{-1}$.
Calibration was done with a standard chopper wheel method, and the
intensities are expected to have a precision better than $\sim$10\%.

The line emission was mapped on grids spaced uniformly by
20\arcsec$\:$ starting at the center of each remnant.  The extents of
the mapped regions were chosen as appropriate to encompass the regions
of bright CO emission discovered near the SNRs.  For most of the
observations the telescope was operated in frequency-switched mode
with a throw of 21 MHz (corresponding to a velocity shift of about 27
km s$^{-1}$), although a subset of positions were observed in
position-switched mode as a consistency check and to search for broad
velocity wings to the CO emission.  We chose the appropriate cloud
velocity at which to observe from the data reported by Cohen \etal\
(1988). The total integration time at each grid position was usually
1800 s, although some shorter exposures were taken near N132D and N23.
The system temperature outside the atmosphere was between 800 K and
1000 K during the two days N49 was observed, 1000 K and 1100 K for the
two days N23 was observed, and varied from 1000 K to 1600 K for N132D.

\subsection{Reduction}

The spectrum at each scan position consists of 2000 channels which
oversample the velocity resolution of the receiver by about a factor
of 2.  The frequency-switched data were reduced in the following
manner.  The spectrum was copied, shifted in velocity by the 21 MHz
throw, inverted to correct for the amplitude inversion between the two
phases, and then averaged with the original spectrum.  The baselines
of the frequency-switched data include temporal variations in the sky
and instrument, and the folding procedure removes most of these
effects.  Only the central 45 km s$^{-1}$ is used during the rest of
the reduction.

The continuum fitting task in IRAF, noao.onedspec.continuum, was used
to fit baselines.  Of the several fit parameters available, our
analysis showed that the most important were (1) the type of function
to fit to the data, (2) its order (generally low to avoid fitting
noise), (3) the number of discrepant points to reject, and (4) the
velocity range of the data over which to fit.

Several combinations of parameters produced final spectra with very
similar quality baseline fits. Nevertheless we were concerned that the
baseline fits could introduce a bias in the derived value of the
velocity integrated CO brightness temperature, $W_{\rm CO}= \Delta v
\sum T_i$, where the summation is over the line profile, $T_i$ is the
observed temperature in channel $i$, and $\Delta v $ is the velocity
width of each channel (which is a constant for our SEST data).  The
set of parameters that we ultimately used for baseline subtraction
were those that gave a zero mean value for $W_{\rm CO}$ in spectra
which appeared to have no signal.  These line-free scans were at the
outer regions of each cloud, where, if there was any signal present,
it was too weak to be distinguished from the noise.  Since the
baselines of all the lines were similar, we are confident that the
functions that behaved well in the line region of scans with no signal
were also well-behaved in scans with prominent CO line emission.

Line centroids and widths were computed directly from the individual
spectra weighting by the observed brightness temperature in each
spectral channel. All velocities are given in the local standard of
rest (LSR) frame. The $W_{\rm CO}$ velocity integration range was kept
the same for all scans of each target. This interval was determined by
examining the location and width of the line in the summed ensemble
spectrum of the cloud. Although there is some velocity gradient within
the clouds, there was not enough to prevent an identical procedure
from being followed for all the scans of a single cloud.

The position-switched scans did not need to be folded, but the same
baseline fitting procedures and error analyses were used.  Lower order
polynomial fits were used because the baselines were flatter.  These
data were used as a check on the frequency-switched scans and to
search for broad velocity components in the detected clouds.

\subsection{Error Analysis}

The primary source of error in $W_{\rm CO}$ is the baseline fitting
procedure.  We estimated the error in our fits by comparing the values
of $W_{\rm CO}$ obtained with two different fitting functions: a
polynomial of the $n$-th order versus one of order $n$+1.  The $W_{\rm
CO}$ values derived from these two fits were correlated and fitted by
least squares to a linear relation.  The root-mean-square (RMS)
deviation of the various points from the linear least squares fit
provided the error estimate, $\delta_{\rm BL}$, from baseline fitting
for an single spectrum.

To quantify the noise in the observations, the antenna temperatures
in velocity channels free of emission were examined.  When these
values were plotted the histogram was Gaussian.  We then sorted these
values and extracted the temperature values corresponding to the
15.87 and 84.13 percentiles in the list.  Half the difference between
these values was taken to be the one $\sigma$ noise estimate per
channel. This error, $\delta_T$, was propagated through the derived
quantities, $W_{\rm CO}$, line centroids, and widths.

In our error analysis we propagate the temperature noise per channel,
$\delta_T$, through to the quantities derived above.  In the case of
$W_{\rm CO}$, the baseline fitting error ($\delta_{\rm BL}$) is
included by direct summation (not root-sum-square) with the noise
error $\delta_{W_{\rm CO}} = \sqrt{N} \Delta v \delta_T + \delta_{\rm
BL}$, where $N$ is the number of velocity channels over which all
summations were done.

\section{Analysis and Results}
\subsection{N49}

The CO emission of the cloud near the N49 mapped by SEST is rather
weak and various different parameters were used in an effort to get
the best baselines.  The function that provided the best fits was a
third-order Legendre polynomial which rejected no points and fit the
regions from $+$263 to $+$280 km s$^{-1}$ and again from $+$292 to
$+$306 km s$^{-1}$.  These regions were chosen so as not to include
the line or the effects present from the folding procedure.
$W_{\rm{CO}}$ was calculated for the velocity region of $+$281 to
$+$291 km s$^{-1}$, as were the line center and width. A summary
of the fit parameters and error analysis is given in Table 1.

To derive values for $W_{\rm CO}$, velocity, and line width for the
ensemble cloud, we produced a composite spectrum by averaging the
baseline-subtracted
scans with individual $W_{\rm CO}$ values that were greater than
3$\delta_{W_{\rm CO}}$ in order to include actual signal and avoid
unnecessary noise.  We plot this in Figure 1a.  As described above,
the optimum baseline fit was chosen for how well it subtracted the
baseline in the line region. However, there remains a small residual
baseline outside the line region due to the observational conditions.
Table 2 lists the information derived from the averaged spectrum,
showing the number of scan positions averaged, the velocity-integrated
CO brightness temperature, the velocity centroid, and the
root-mean-square velocity width of the line.

The mean $W_{\rm CO}$ of the
``zero'' scans was 0.09${\pm}$0.32 K km s$^{-1}$.  This error is
comparable to the $\delta_{W_{\rm CO}}$ derived independently using
the techniques outlined above.  Figure 2 shows the individual spectra
in their correct relative positions on the sky.

Two position-switched spectra were taken at the location of the peak
CO emission. The derived $W_{\rm CO}$ values, line centroids, and
widths were consistent with those from the frequency-switched data.
No evidence for a broader component to the line was apparent.

\subsection{N132D}

The same reduction technique was used for N132D.  The function that
gave the best baseline fit to the N132D data was a fifth-order
Legendre polynomial with no rejection, and fit over the regions from
$+$243 to $+$256 km s$^{-1}$ and $+$272 to $+$286 km s$^{-1}$.
$W_{\rm{CO}}$ and velocity information were calculated in the region
from $+$257 to $+$271 km s$^{-1}$. The velocity window for the line
emission was wider here than for N49 because of a significant velocity
gradient ($\sim$5 km s$^{-1}$) across the N132D cloud.  The mean
$W_{\rm{CO}}$ of the ``zero'' scans was $-0.070\pm 1.27$.

The error analysis for N132D was broken down into two parts, which are
listed separately in Table 1.  Upon examination of the data, it became
obvious that there was much greater noise in the scans taken on
October 17 and 19, 1990, which was likely the result of poorer weather
conditions.  The noise and fit statistics for these days were
considered separately from the rest of the data.  The total error in
$W_{\rm CO}$ (as the sum of the baseline and noise errors) was used in
determining which scans were to be averaged to the total cloud
spectra.  Therefore, by considering the data in two separate sets,
only true signal was added because the noisier data had a higher
3$\delta_{W_{\rm CO}}$ threshold.

Once again, to derive values of $W_{\rm{CO}}$, velocity, and line
width for the whole cloud, baseline-subtracted spectra with signal
greater than 3$\delta_{W_{\rm CO}}$ were averaged.  Table 2 lists
information derived from this spectrum and the spectrum itself is
shown in Figure 1b. Figure 3 plots the individual spectra throughout
the cloud.

We took some position-switched data at five positions located near the
southern X-ray emitting rim of N132D in order to search for a broad
component to the line which might indicate interaction between the
cloud and N132D.  These data were reduced as described above, but with
linear baselines.  The same velocity ranges as for the
frequency-switched data were used for the baseline fits and to
calculate the line values. The mean difference in $W_{\rm CO}$ when
compared with the frequency-switched scans at the same positions was
0.6 K km s$^{-1}$, within $\delta_{W_{\rm{CO}}}$.  Because the
baselines are flatter than the frequency-switched data, it was
possible to search for broad velocity components to the main line, as
well as other weaker lines over the entire velocity region from $+$225
to $+$335 km s$^{-1}$.  We began by summing all the position-switched
scans to form a high signal-to-noise spectrum.  This spectrum was fitted
initially with a single Gaussian to quantify the main peak.  To check
for a broad CO emission line, we then added a broad Gaussian (with a
fixed FWHM of 35 km s$^{-1}$) centered on the line region to the main
line.  The reduction in $\chi^2$ when this broad line was included was
not statistically significant. There were also weak lines at $+$243
and $+$285 km s$^{-1}$ that also turned out not to be statistically
significant. In conclusion, our data provide no evidence for either a
broad component to the main line nor for additional lines beyond that
seen in the frequency-switched data.

\subsection{N23}

The frequency-switched scans taken near N23 were reduced with
fifth-order Legendre polynomials with no rejection of discrepant
points.  The baseline was fit from $+$243 to $+$274 km s$^{-1}$ and
from $+$286 to $+$306 km s$^{-1}$.  The position-switched data were
reduced with a second order Legendre polynomial baseline subtraction
and fit from $+$225 to $+$274 km s$^{-1}$ and from $+$286 to $+$335 km
s$^{-1}$.  Figure 4 shows the spectra taken around the remnant in
their positions on the sky.

Although the scan pattern around N23 may appear haphazard, there is an
explanation for the apparently odd placement.  The position-switched
scans (Figure 4), which were carried out first, were arranged in a
pattern similar to those for N132D and N49 -- centered on the remnant
and extending toward the edge where the X-ray emission appears
strongest.  When no CO emission was found there and, since the
observing time was running short, the mapping strategy was changed in
an attempt merely to locate the CO emission, which according to the
Cohen \etal\ (1988) map, should have extended generally toward the
northwest. Two areas were mapped: a $9\myarcmin.3$ long north-south
strip of scans and a 2\arcmin$\:$ square pattern to the
northeast. These spectra were integrated for only 120 s and were taken
in frequency-switched mode. None of these scans show significant CO
emission, as indicated in Table 3.

\subsection{X-ray Images}

For comparison with our maps of CO emission, we obtained high
resolution X-ray images from the \rosat\ and \einstein\ archives.  N49
was observed by the \rosat\ high resolution imager (RHRI) for a total
live-time corrected exposure of 41972.4 s in several intervals from
March 1992 to March 1993 (ROR numbers 400066 and 500172).  N132D was
observed by the RHRI in February 1991 for 26830.8 s (ROR number
500002). Since the RHRI observations of N23 are not yet publicly
available, we used the \einstein\ HRI data instead. The EHRI observed
N23 in May 1980 for 15433.3 s. 

All images were deconvolved with a small number of iterations of the
Lucy-Richardson algorithm using the implementation in IRAF. The
resulting X-ray maps are shown in Figures 5, 6, and 7 for N49, N132D,
and N23, with the SEST scan positions and contours of $W_{\rm CO}$
emission overlaid.  Our X-ray maps of N49 and N132D agree well with those
published by Mathewson \etal\ (1983); however our N23 map shows much
less structure than the previously published one, although the overall
appearance is similar, since our effective smoothing of the X-ray
image is greater than the 2$^{\prime\prime}$ Gaussian sigma used
before.

\section{Discussion}
\subsection{N49}

The peak X-ray and optical emission from N49 lies along the eastern
limb and coincides extremely well with the position of the mapped CO
emission (Fig.~5). In addition to the agreement in projected position,
the agreement in velocity, or line-of-sight position, is also quite
good.  N49 shows a narrow emission line in its optical spectrum that
arises from the photoionization of unshocked gas at rest with respect
to the local environment and preceding the supernova blast wave
(Shull 1983, Vancura \etal\ 1992a).  The LSR velocity of this material
is $+$286$\pm$1 km s$^{-1}$ which is in excellent agreement with our
integrated cloud velocity of $+$286.0$\pm$0.1 km s$^{-1}$.

The prevailing picture of the N49 environment is one of relatively
high density with a gradient in the ambient density increasing from NW
to SE. Vancura \etal\ (1992a) estimate the mean preshock density of the
intercloud medium surrounding N49 to be 0.9 cm$^{-3}$ based on the
observed X-ray emission. They find that the optical emission must
arise from much denser regions with a range of preshock densities
covering 20 to 940 cm$^{-3}$.  The lack of spectral variations with
brightness can be well explained in terms of sheets of optical
emission formed as the supernova blast wave encounters a large dense
cloud of gas. The molecular cloud that we have discovered provides a
natural explanation for this general picture.

N49 lies at the northern edge of a complex infrared emitting region.
The closest cataloged IRAS source (from the Leiden--IRAS Magellanic
Clouds Infrared Source Catalogues, Schwering \& Israel 1990), LI-LMC
1022, lies just north of the CO cloud at a position of 05:25:59.5,
$-$66:07:03 (B1950). Graham \etal\ (1987) have argued that
collisionally-heated dust in the SNR could be the explanation for the
IRAS source. We consider it unlikely that this is the entire
explanation, since the flux of LI-LMC 1022 (19.5 Jy at 60 $\mu$m)
corresponds to a luminosity about an order of magnitude larger than
the far infrared luminosity of similar sized Galactic remnants where
comparative morphology and other considerations make it clear that we
are observing heated dust in the SNR (Saken, Fesen, \& Shull 1992).
In addition, the four times higher gas-to-dust ratio of the LMC
(Koornneef 1984) would suggest lower comparative IR fluxes.

It is more likely that the molecular cloud is the origin of the IR
emission.  Israel \etal\ (1993) detect CO emission from a large
fraction (87\%) of a sample of LMC IRAS sources over a wide range of
infrared luminosity.  Far infrared emission can arise from the heating
of dust in a molecular cloud by embedded stars, the general
interstellar radiation field, stellar radiation from nearby star
clusters, or other sources.  Caldwell \& Kutner (1996) have studied a
number of molecular clouds in the LMC using the ratio of far infrared
luminosity $L_{\rm FIR}$ to the cloud virial mass $M_{\rm V}$ as a
measure of star formation activity, assuming that the infrared
luminosity arises from embedded young stars. For LI-LMC 1022 we
estimate $L_{\rm FIR} \sim 13\times 10^4\, L_\odot$, while the cloud
mass is $M_{\rm V} \sim 3\times 10^4\, M_\odot$ (see below).  The
derived ratio $\sim$4 $L_\odot/M_\odot$ is well within the range of
other LMC clouds studied by Caldwell \& Kutner.

The remnant N49 is itself a strong source of optical, UV, and
X-radiation of which Ly$\alpha$ and the O {\sc vi} $\lambda$1035
doublet are the principal sources of flux. Our estimate of N49's
intrinsic Ly$\alpha$ luminosity, $2.1\times10^{38}$ ergs s$^{-1}$, is
based on a two-photon continuum flux of $2.3\times 10^{-10}$ ergs
cm$^{-2}$ s$^{-1}$ (Vancura \etal~1992a) and a value of 3 for the
ratio of Ly$\alpha$ to two-photon continuum, as expected from models
of planar shocks over a broad range in velocity (Hartigan, Raymond, \&
Hartmann 1987).  We estimate the O {\sc vi} $\lambda$1035 luminosity
by scaling C {\sc iv} $\lambda$1550 ($2.8\times10^{37}$ ergs s$^{-1}$)
by a factor 4 (Vancura \etal~1992b).  Together with the X-ray
luminosity of $1.9\times10^{37}$ ergs s$^{-1}$, this yields a total
luminosity of hard photons of $3.4\times 10^{38}$ ergs s$^{-1}$.
If the molecular cloud intercepts half of these photons and all that
luminosity is subsequently re-radiated in the infrared band, then we
would expect $L_{\rm FIR} \sim 4\times10^4\, L_\odot$, which is
potentially a sizeable fraction, about one-third, of that actually
seen.

Since there are no other obvious signs of active star formation
(cataloged OB associations or \hii\ regions) in the vicinity of N49,
heating by the interstellar radiation field may be negligible and an
embedded source origin for the remainder of the far IR emission may
need to be considered.  However, since this region of the LMC is
rather complex and the angular resolution of IRAS is modest
($\sim$1$^\prime$), a definitive explanation for the origin of this
emission awaits a more comprehensive study of the N49 environment
using higher angular resolution IR data.

\subsection{N132D}

Figure 6 shows the spatial relationship between N132D and the bright
CO cloud discovered near it. To complete the association between them,
we turn to the photoionization precursor in the quiescent gas upstream
of the expanding supernova blast wave that N132D (like N49) displays.
Morse, Winkler, \& Kirshner (1995) studied this component to the
optical emission and found it to display apparently normal LMC
abundances and to be spectrally unresolved at a resolution of 30 km
s$^{-1}$. Its LSR velocity of $+268\pm7$ km s$^{-1}$ is in excellent
agreement with our velocity of $+$264.0$\pm$0.1 km s$^{-1}$ for the
cloud. As with N49, this agreement of velocities along with the
proximity in projected position implies a definite physical
association.

Based on a study of its X-ray morphology, Hughes (1987) proposed that
N132D was expanding into a region with a density gradient increasing
from northeast to southwest.  To explain the X-ray emission, mean
preshock densities of 2--3 cm$^{-3}$ were required in the southern
(denser) region. Recently Morse \etal\ (1996) studied the optical
photoionization precursor in this area in more detail and derived a
preshock density of roughly 3 cm$^{-3}$ from the surface brightness of
[O {\sc iii}] $\lambda$5007.  The presence of a dense molecular cloud
toward the south of the remnant provides a general framework for
understanding these results.  N132D's incomplete morphology (i.e., the
lack of emission or ``break-out'' to the northeast), the fact that the
pre-shock ambient density is considerably lower than the densities
usually associated with the cores of molecular clouds, and the spatial
separation between the bright CO core and the remnant itself, strongly
suggest that N132D lies near the northern boundary of its associated
molecular cloud.  Numerical hydrodynamic simulations of the explosion
of a SN near the edge of a molecular cloud (Tenorio-Tagle,
Bodenheimer, \& Yorke 1985) do bear some similarity to the observed
features of N132D. Given the wealth of specific information known
about N132D, further numerical work in this area could yield important
information on the structure of molecular clouds and their interface
with the general ISM.

Morse \etal\ (1995) first pointed out an apparent association between
the CO emission (as published by Hughes \etal\ 1991) and an \hii\
region about 2$^\prime$ south of N132D.  Figure 8 shows an optical
H$\alpha$ image of both the SNR and \hii\ region overlaid with
contours of CO emission from our current analysis.  The position and
shape of the \hii\ region agree remarkably well with the cloud core.
There is also an IRAS source nearby at 5:25:32.3, $-$69:43:28 (B1950)
(LI-LMC 1008, Schwering \& Israel 1990) with a flux at 60 $\mu$m of
24.8 Jy. The inferred far infrared luminosity $L_{\rm FIR} \sim
16\times 10^4\, L_\odot$ compared to the virial mass of the cloud (see
below) $M_{\rm V} \sim 2\times 10^5\, M_{\odot}$ yields a ratio
$\sim$1 $L_\odot/M_\odot$ which is consistent with other LMC clouds
(Caldwell \& Kutner 1996) and indicates a substantial amount of
heating of the cloud. Like N49, N132D has a substantial UV and X-ray
flux.  Integrating N132D's effective ionizing spectrum (Morse \etal\
1996), we estimate the remnant's intrinsic luminosity (including
Ly$\alpha$ and harder photons) to be $1.8\times 10^{38}$ ergs
s$^{-1}$. Again assuming that half these photons are absorbed and
re-radiated by the molecular cloud gives us an estimate of the far
infrared luminosity of $\sim$$2\times 10^4\, L_\odot$, which is
evidently only about 10\% of the observed $L_{\rm FIR}$. Another
source of cloud heating is the ionizing flux from the same star (or
set of stars) that is exciting the \hii\ region.

There is a velocity shear of about 5 km s$^{-1}$ across the cloud with
the peak of the emission increasing from $\sim$263 km s$^{-1}$ in the
east to $\sim$268 km s$^{-1}$ in the west (see Fig.~3).  In addition
to this velocity shear, there are asymmetric wings on the line
profiles that extend to higher velocities in the eastern part of the
cloud and to lower velocities in the west. These results are
qualitatively consistent with simple rotation of the cloud (Dubinski,
Narayan, \& Phillips 1995).  However, we favor a somewhat different
interpretation. 
We believe that there are (predominantly) two clouds at radial
velocities of $\sim$263 km s$^{-1}$ and $\sim$268 km s$^{-1}$, with
intrinsic velocity widths $\sigma \sim 1.8$ km s$^{-1}$, containing
bright dense cores centered near the CO peaks in the east and west of
our maps.  These cores are each surrounded by lower density material
covering larger regions of the sky.  The emission we observe at any
particular position in the map is a blend of these two components.
For example, consider the set of spectra at declination $-$69:42:22
(corresponding to the fourth row of scan points south of position 0,0 in
figure 3). At right ascension 5:25:27 (immediately below position 0,0)
and 5:25:23 (one scan position to the west of 0,0) we see emission that is
considerably broader and weaker than scans further to the east or
west.  In these positions, which lie between the brighter cores, we
suggest that we are seeing nearly equal contributions from the lower
density ``halos'' of the two separate clouds. Further to the east or
west the  spectra are dominated by a single component, but the weaker
emission from the other cloud remains as an asymmetric wing on the
line profile. In any event, our data 
reveal that this region is quite complex both morphologically and
dynamically.

\subsection{N23}

In general N23 is not as well studied as the previous two remnants.
Its X-ray morphology (Fig.~7) is similar to that of N49, showing a
brightening toward the east.  In this same direction the optical image
(Mathewson \etal\ 1983) also shows a brightened limb and, in addition,
there appears to be a cluster of bright stars beyond the supernova
shock front.  The nearest cataloged IRAS source is some
200$^{\prime\prime}$ south of N23.  Two other IRAS sources lie
4$^\prime$ and 5$^\prime$ west and southwest.

As mentioned above, the remnant sits near the southern boundary of a
modest CO cloud (\# 9 from Cohen \etal). However as our results on
N49 and N132D clearly show, these lower angular resolution data
provide only a coarse guide to the existence of molecular gas on
sub-arcminute spatial scales. We were unable to map a large region
near N23 to a sensitive level (due to limited observation time) and so
our null result on the association of a molecular cloud with this
remnant must be considered tentative.  We point out that our upper
limit to CO emission over the region surveyed northeast of the remnant
is consistent with the velocity-integrated CO emission from the cloud
actually detected near N49.  This region would be an interesting one
to follow-up with additional SEST observations.

\section{Cloud Masses}

The mass of a spherical, self-gravitating molecular cloud of radius
$R$ in virial equilibrium can be estimated as $M_{\rm V} =
5R\sigma^2/{\rm G}$, where $\sigma$ is the line width (RMS) of the ensemble
cloud spectrum.  We determine the effective radii of the clouds from
our maps as $R = (A/\pi)^{1/2}$, where the cloud area $A$ is given by
$A = N_S\,L^2$. $N_S$ is the number of scans with significant signal
averaged in the ensemble
spectrum (see Table 1) and $L$ is the spacing between our scans
(20\arcsec$\:$ or 4.85 pc at the LMC). The effective cloud radii are
7.2 pc (N49) and 22.9 pc (N132D).  For the N49 cloud we derive a
virial mass of $M_{\rm V} = 3\times 10^4\, M_{\odot}$ and for the
N132D cloud we find $M_{\rm V} = 2\times 10^5\, M_{\odot}$.

It is also possible to estimate molecular cloud masses using the
empirical relationship between velocity-integrated CO intensity and
H$_2$ column density. A recent redetermination of the ratio between
these quantities for the \jonezero\ transition of CO in the Galaxy
yields a value $X_G = N_{\rm H_2} / W_{\rm CO(1-0)} = 1.56\times
10^{20}$ cm$^{-2}$ K$^{-1}$ km$^{-1}$ s (Hunter \etal\ 1996).  For our
data on clouds in the LMC, we must correct this factor for (1) the
observational fact that LMC molecular clouds are intrinsically less
luminous in CO than Galactic clouds of similar mass (due in part to
the lower metallicity and the higher gas-to-dust ratio of the LMC),
and (2) the intensity ratio between the \jtwoone\ and the \jonezero\
transitions of CO. We use the scaling given by Cohen \etal\ (1988),
$X_{\rm LMC} = 6 X_{\rm G}$ to account for the first item.
Sakamoto \etal\ (1995), based on a study of molecular clouds in the
first quadrant of the Galaxy, find a value of $0.66$ for the mean
ratio $W_{\rm CO(2-1)}/W_{\rm CO(1-0)}$, with a variation from 0.5 to
0.8 as a function of Galactocentric distance.  This value is in
substantial agreement with other recent measures of this ratio, such
as that of Chiar \etal\ (1994), who find a value of $0.85\pm0.63$ from
a study of molecular clouds in the Scutum arm of the Galaxy.  For
clouds in the Chiar \etal\ study with $M_{\rm V} < 
5 \times10^4 \, M_{\odot}$ based on the \jtwoone\ transition,
we find a ratio of $\sim$0.5, which should be the appropriate
value to use for the N49 cloud. For clouds with $M_{\rm V}$ in the
range (1.58 -- 2.32) $\times10^5 \, M_{\odot}$ and therefore similar
to the cloud near N132D, the ratio is slightly higher, $\sim$0.8.
Combining these factors, we come up with a
relationship of $N_{\rm H_2}/W_{\rm CO(2-1)} = 1.87 (1.17) \times
10^{21}$ cm$^{-2}$ K$^{-1}$ km$^{-1}$ s for the N49 (N132D) cloud.
The masses we derive (including a correction for He assuming [He]/[H]
= 0.085) are $M_{\rm CO} = 9\times 10^3\, M_\odot$ (N49 cloud) and
$M_{\rm CO} = 3\times 10^5\, M_\odot$ (N132D cloud).

The giant molecular cloud near N132D shows good agreement
between its CO-derived mass and the virial mass. It also
represents a significant fraction, roughly 30\%, of the total CO mass
of the cloud complex near N132D from the lower resolution Cohen \etal\
(1988) complete survey of the LMC (cloud \# 22 in their Table 1).
This complex has an estimated mass of $9.5\times 10^5\, M_\odot$ after
including the revised $X_G$ factor from above.  The smaller molecular
cloud we found near N49 is far from being a major constituent of the
corresponding cloud (\# 23) in the LMC survey, encompassing less than
0.5\% of the total mass.  In addition, the SEST cloud may not 
be in virial equilibrium based on the considerable difference between
its CO mass and virial mass, the latter being over three times larger
than the former.  The use of this observation as evidence, albeit
weak, for a direct interaction between the cloud and N49's blast wave
must be tempered by strong caveats about the large uncertainties in
these mass estimates.  A convincing case for interaction between these
SNRs and their associated clouds will require new considerably deeper
observations and the discovery of broad molecular emission lines
(30--40 km s$^{-1}$) from the interaction region.

\section{Summary}

We used the SEST to map the vicinity of three LMC SNRs, N49, N132D,
and N23, in the \jtwoone\ transition of CO emission and found that:

(1) The SNRs N49 and N132D show spatial relationships with molecular
clouds that coincide with increased X-ray and optical emission from
the remnants.  There is also good agreement between the mean velocity
of the ensemble CO emission, $+$$286.0\pm0.1$ km s$^{-1}$ (N49) and
$+$$264.0\pm0.1$ km s$^{-1}$ (N132D), and the optically determined
velocities of the remnants, $+$$286\pm1$ km s$^{-1}$ (N49, Shull 1983)
and $+$$268\pm7$ km s$^{-1}$ (N132D, Morse \etal\ 1995). The agreement
of the SNR and cloud velocities along with the two-dimensional
proximity indicate that the two systems are indeed physically
associated.

(2) CO and virial equilibrium masses were derived for the two newly
discovered molecular clouds.  The different mass estimates agree quite
well for the N132D cloud and indicate a mass of $\sim$$3\times 10^5\,
M_\odot$ which is within the range expected for a giant molecular
cloud.  Our two mass estimates for the cloud near N49 are internally
inconsistent, which may indicate that the cloud is not virialized.
The CO mass for this cloud is $9\times 10^3\, M_\odot$ while the
virial mass is $3\times 10^4\, M_\odot$. In neither
case does the SEST cloud account for the entire emission observed at
lower spatial resolution by Cohen \etal\ (1988) in their complete LMC
CO survey.  The N49 SEST cloud is indeed a negligible fraction
($<$0.5\%), although the N132D SEST cloud corresponds to about
30\% of the total cloud mass observed at lower resolution.

(3) We detected no CO emission from the vicinity of the SNR N23.
Since the area mapped was limited, this null result should not be
overinterpreted. The X-ray and optical morphology of N23 and its
proximity to CO emission in the Cohen \etal\ (1988) survey continue to
support the presence of dense molecular gas near N23 and further
observations with the SEST are warranted.

(4) The association of N49 and N132D with dense molecular clouds
supports the picture in which their progenitors were short-lived,
hence massive, stars that exploded as core collapse (i.e., Type II or
Type Ib) supernovae.  This is consistent with other known
characteristics of the remnants, such as the presence of
high-velocity, oxygen-rich stellar material in N132D, which is the
traditional signature of a massive star progenitor.  The work we
present here has also made clear that the good angular resolution of
the SEST provides the essential key for being able to make
SNR/molecular cloud associations in the LMC. Discovery of CO emission
in the near vicinities of other LMC SNRs with the SEST would allow us
to determine the SN type of a larger fraction of the remnant sample in
the Cloud and should be pursued.

\vspace{0.5in}

The Swedish-ESO Submillimeter Telescope, SEST, is operated jointly by
ESO and the Swedish National Facility for Radio Astronomy, Onsala
Space Observatory at Chalmers University of Technology.  This research
has made use of data obtained through the High Energy Astrophysics
Science Archive Research Center Online Service, provided by the
NASA-Goddard Space Flight Center.  We thank Jon Morse for providing
the H$\alpha$ image of N132D and the nearby \hii\ region and we
acknowledge very useful discussions with Mark Birkinshaw, Tom Dame,
and John Raymond.  We would also like to thank Kristin Kearns, Kim
Dow, and everyone associated with the Smithsonian Astrophysical
Observatory Summer Intern Program, which was funded through the
National Science Foundation.  L.B. acknowledges support by FONDECYT
Grant 1950627, Rep\'ublica de Chile. Additional financial support for
this research was provided by NASA (\rosat\ Grant NAG5-2156) and the
Smithsonian Institution.

\clearpage
\begin{deluxetable}{cccccc}
\tablecaption{Error analysis of CO line emission}
\tablewidth{0pt}
\tablehead{
\colhead{Associated} & \colhead{Baseline Fit Region} &
\colhead{Line Region} & \colhead{$\delta_T$} 
&\colhead{$\sqrt{N} \Delta v \delta_T$}  & \colhead{$\delta_{BL}$} \nl
\colhead{SNR} &\colhead{ (km s$^{-1}$)}
&\colhead{(km s$^{-1}$)} & \colhead{(K)}
&\colhead{ (K km s$^{-1}$)} &\colhead{ (K km s$^{-1}$)}  }
\startdata
N49 & 263$\,$--$\,$280,292$\,$--$\,$306 & 281$\,$--$\,$291 & 0.13 &
0.093 & 0.22\nl
\nl
N132D (total) & 243$\,$--$\,$256,272$\,$--$\,$286 & 
257$\,$--$\,$271 & 0.32 & 0.28 & 0.37\nl 
N132D (high noise) & 243$\,$--$\,$256,272$\,$--$\,$286 &
257$\,$--$\,$271 & 0.58 & 0.52 & 0.52 \nl
N132D (low noise) & 243$\,$--$\,$256,272$\,$--$\,$286 &
257$\,$--$\,$271 & 0.22 & 0.20 & 0.18 \nl
\nl
N23 (pos switched)& 225$\,$--$\,$274,286$\,$--$\,$335 &
275$\,$--$\,$285 & 0.23 & 0.17 & 0.30 \nl
N23 (freq switched)& 243$\,$--$\,$274,286$\,$--$\,$306 &
275$\,$--$\,$285 & 0.70 & 0.53 & 1.5 \nl
\enddata
\end{deluxetable}

%\clearpage

\singlespace
\begin{deluxetable}{ccccc}
\tablecaption{CO line emission for molecular clouds near N49 and N132D}
\tablewidth{0pt}
\tablehead{
\colhead {Associated} & \colhead{ Number of} & \colhead { $<W_{\rm{CO}}>$} & \colhead {$v$} & \colhead {$\sigma$}\nl
\colhead{SNR}& \colhead{Averaged Spectra} & \colhead{(K km s$^{-1}$)}&
\colhead{ (km s$^{-1}$)} & \colhead {(km s$^{-1}$)}}
\startdata
N49 & 7 & $1.72{\pm}0.12$ & $+286.0{\pm}0.1$ & $2.0{\pm}0.1$\nl
N132D & 70 & $9.95{\pm}0.08$ & $+264.0{\pm}0.1$ &
$2.8{\pm}0.1$\nl
\enddata
\end{deluxetable}

%\clearpage

\begin{deluxetable}{ccc}
\tablecaption{CO line emission near N23 (Scan positions explained in
text)}
\tablewidth{0pt}
\tablehead{
\colhead{Scan} &\colhead{ Number of} &\colhead{ $<W_{\rm CO}>$}\nl 
\colhead{Position} &\colhead{Averaged Scans} & \colhead{(K km s$^{-1}$)}}
\startdata
 East of N23 & 9 & $0.01\pm 0.16$ \nl
 Square NE of N23 & 16 &$1.58\pm 0.52$\nl
 Strip North of N23 & 15 & $0.25\pm 0.53$\nl
\enddata
\end{deluxetable}

\clearpage

\figcaption[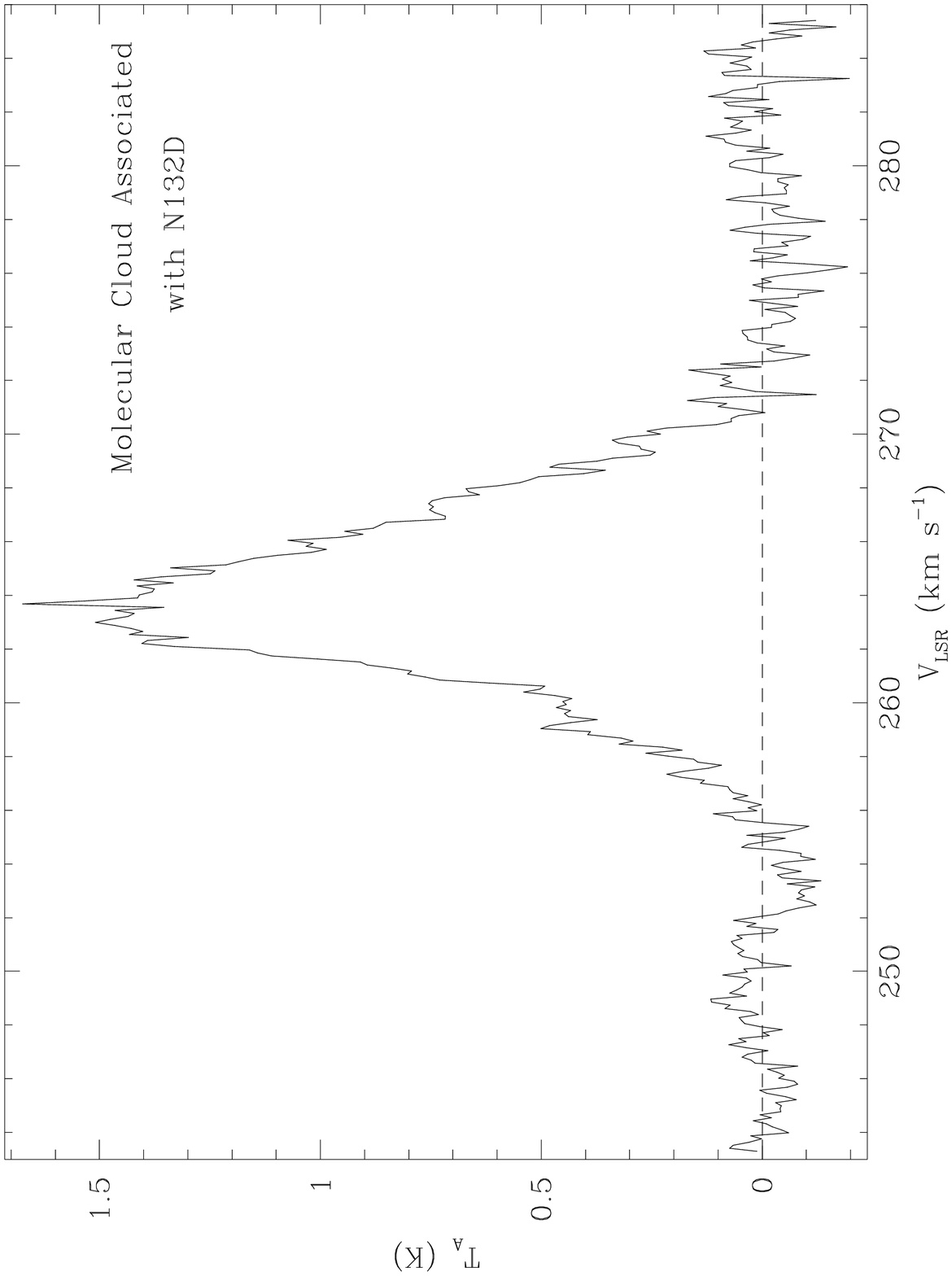]{Ensemble spectra of the molecular clouds near N49 and
N132D.  Only scans with signal stronger than 3$\delta_{W_{\rm CO}}$
are averaged in these spectra.  The zero baseline level is shown.  The
values listed in Table 1 are calculated from these
spectra. \label{fig1}}

\figcaption[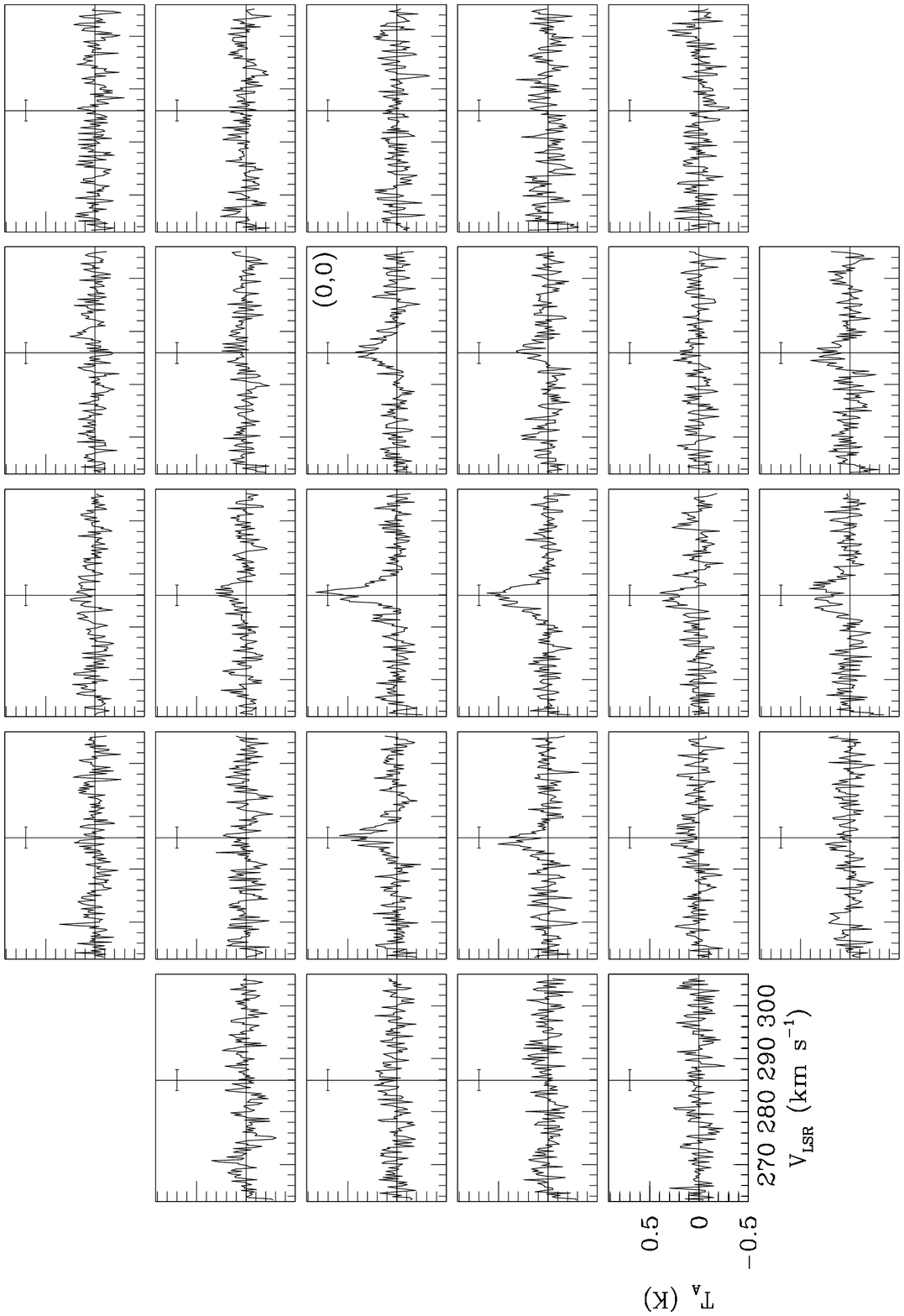]{Spectra of CO \jtwoone\ emission from the
molecular cloud near N49.  The spectra were observed on a grid spaced
20\arcsec$\:$ apart.  The center of N49 is marked as position (0,0).
The reference lines show the zero baseline level, the line center
($+$286.0 km s$^{-1}$) and FWHM (4.65 km s$^{-1}$) of the ensemble
cloud spectra (shown in Figure 1).  (North is to the top, east is to
the left) \label{fig2}}

\figcaption[b.sm11.ps]{Spectra of CO \jtwoone\ emission from the
molecular cloud near N132D.  The spectra were observed on a grid
spaced 20\arcsec$\:$ apart.  The center of N132D is marked as position
(0,0).  The reference lines show the zero baseline level, the line
center ($+$264.0 km s$^{-1}$) and FWHM (6.64 km s$^{-1}$) of the
ensemble cloud spectra (shown in Figure 1).  (North is at the left of
the page and east is toward the bottom.) \label{fig3}}

\figcaption[g.sm5.ps]{Spectra of position-switched CO \jtwoone\
emission near N23.  The center of the remnant is marked with (0,0).
The spectra were observed on a grid spaced 20\arcsec$\:$ apart.  Boxed
comments show the locations of the southernmost positions of short,
frequency-switched scans.  (North is at top, east is to the left)
\label{fig4}}

\figcaption[n49.good3.ps]{\rosat\ X-ray image of N49 overlaid
with velocity integrated CO emission contours.  The X-ray contours are
at 0.35, 0.87, 1.4, 1.9, 2.4, 3.0, and 3.5 counts s$^{-1}$
arcmin$^{-2}$.  The CO contours are 1.0, 1.3, 1.7, 2.0, 2.3, and 2.7 K
km s$^{-1}$.  The grid pattern for the CO observations is
shown. Coordinates are in epoch B1950.
\label{fig5}}

\figcaption[n132d.good5.ps]{\rosat\ X-ray image of N132D overlaid
with velocity integrated emission contours.  The X-ray contours are at
0.54, 1.1, 1.6, 2.1, 2.7, 3.2, and 3.8 counts s$^{-1}$ arcmin$^{-2}$.
The CO contours are 5.0, 8.3, 11.7, 15.0, 18.3, 21.7, 25.0, 28.3, and
31.7 K km s$^{-1}$.  The grid pattern for the CO observations is
shown. Coordinates are in epoch B1950.  \label{fig6}}

\figcaption[n23.good.ps]{\einstein\ X-ray image of N23.  The
X-ray contours are at 0.023, 0.070, 0.12, 0.16, 0.21, 0.26, and 0.30
counts s$^{-1}$ arcmin$^{-2}$.  The grid pattern for the CO
observations is shown.  Coordinates are in epoch B1950. (Squares are
840 second position-switched scans and triangles are short 120 second
frequency-switched scans.) 
\label{fig7}}

\figcaption[n132d_ha_mc.ps]{Narrow-band H$\alpha$ optical image of 
N132D and a nearby \hii\ region from the Rutgers/CTIO imaging
Fabry-Perot spectrometer (Morse \etal\ 1995) overlaid with contours of
molecular emission.  Note the strong spatial correlation between the
molecular cloud and the \hii\ region.  North is at top, east is to the
left. \label{fig8}}

\clearpage
\begin{figure}
\plotfiddle{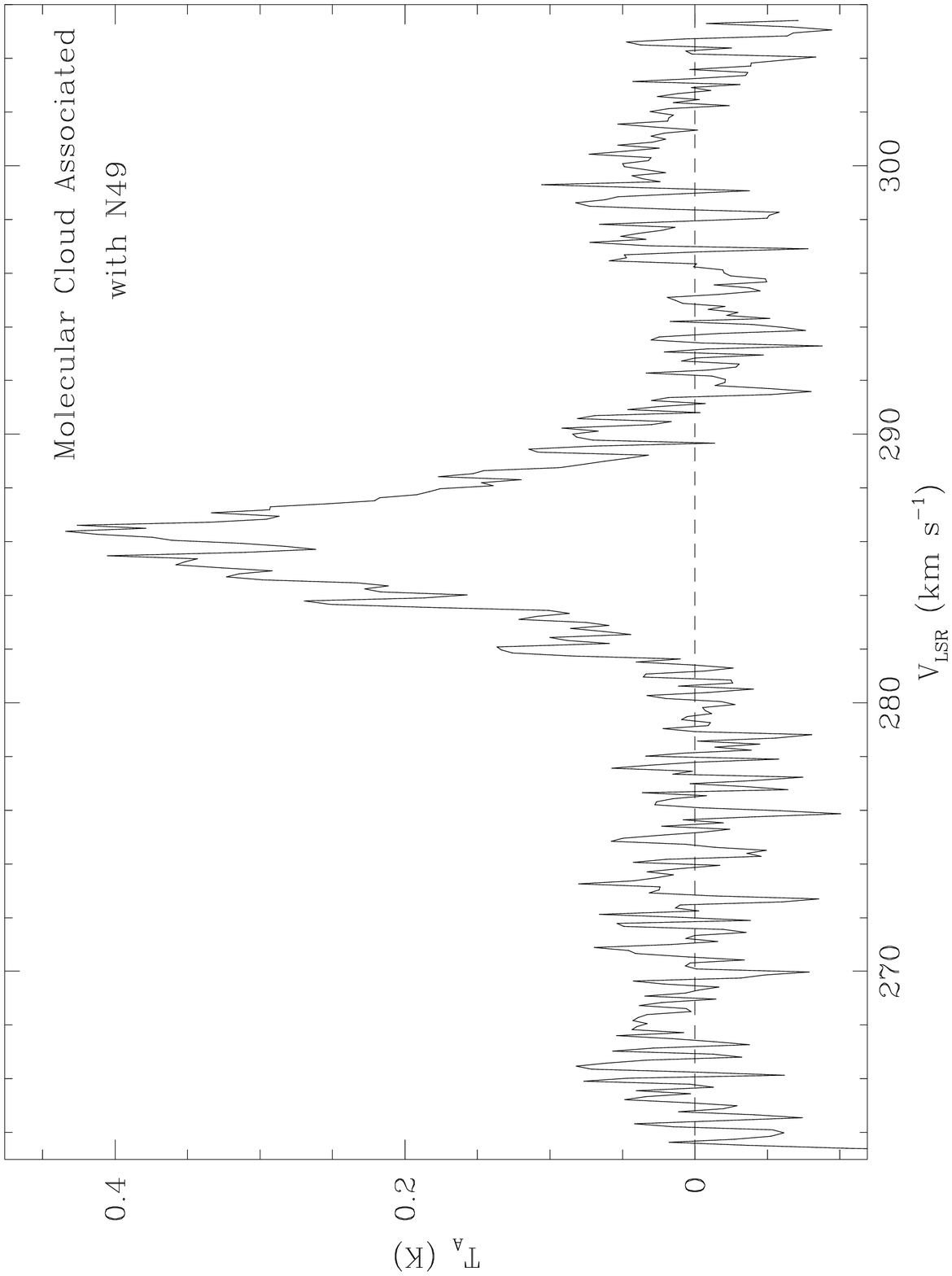}{2.8in}{270}{50}{54}{-200}{290}
\plotfiddle{fig1b.ps}{2.8in}{270}{50}{54}{-200}{220}
\end{figure}

\clearpage
\begin{figure}
\plotfiddle{fig2.ps}{4in}{270}{65}{60}{-285}{350}
\end{figure}

\clearpage
\begin{figure}
\plotfiddle{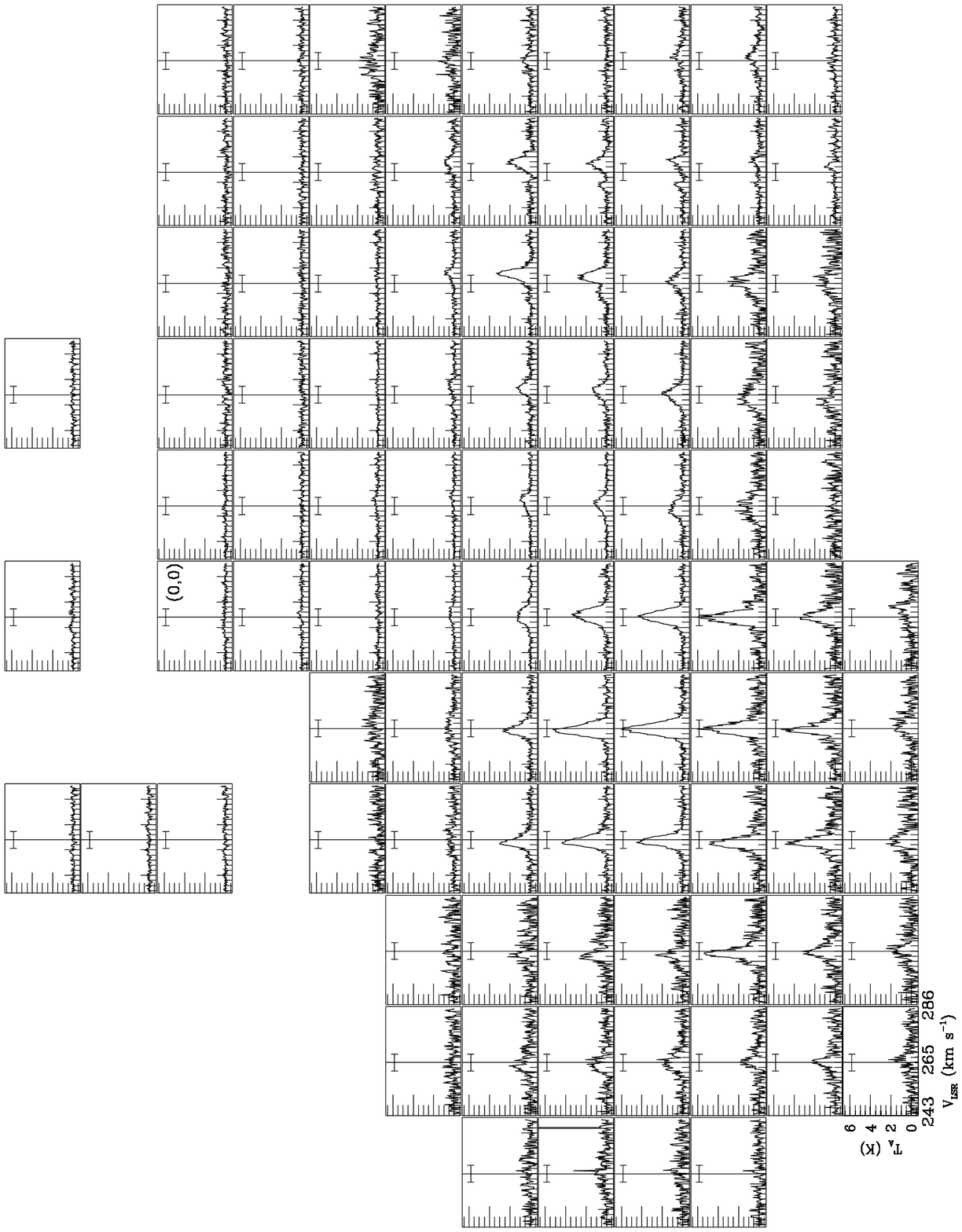}{7in}{0}{100}{85}{-320}{-70}
\end{figure}

\clearpage
\begin{figure}
\plotfiddle{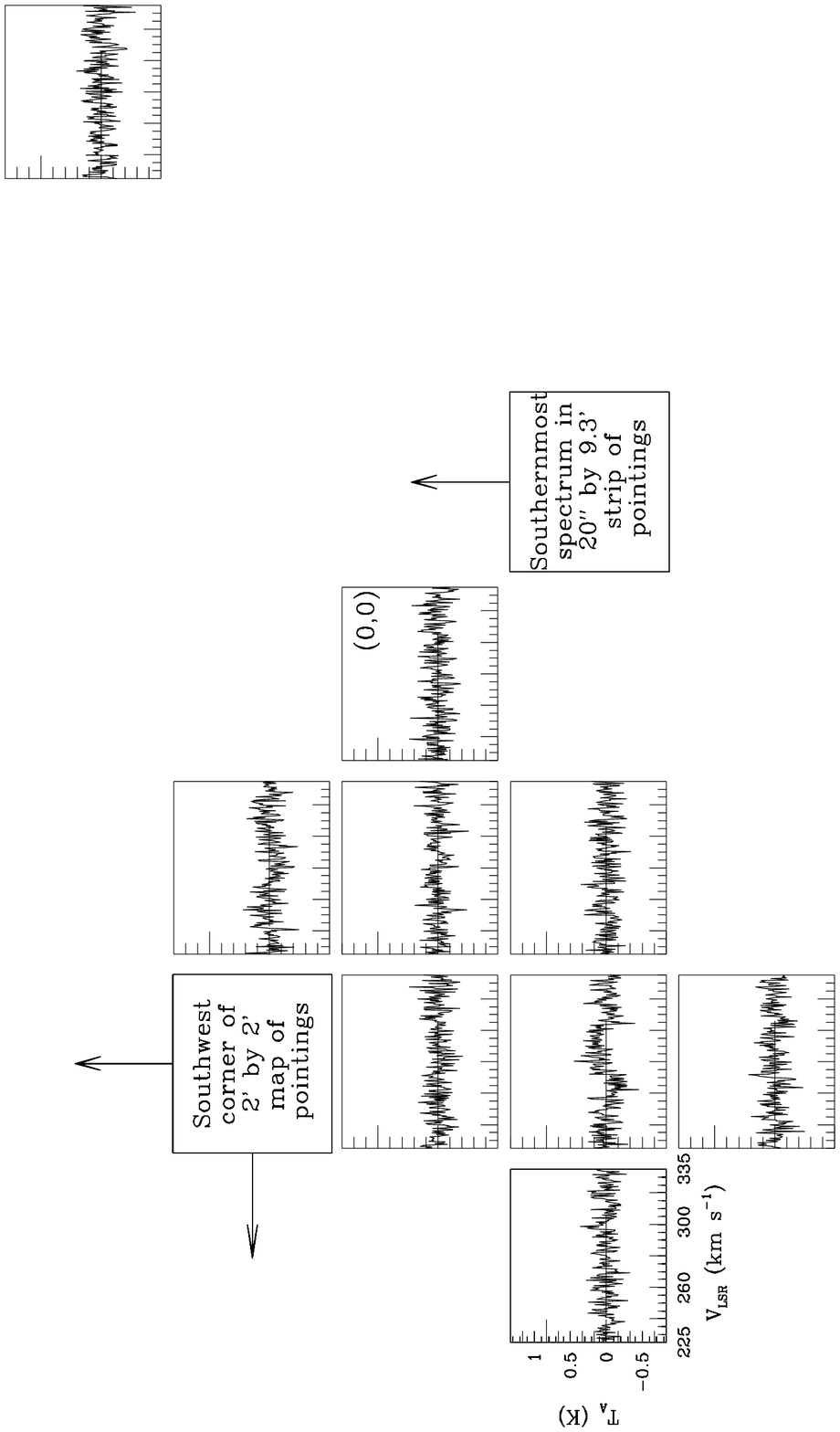}{3in}{270}{65}{60}{-270}{330}
\end{figure}

\clearpage
\begin{figure}
\plotfiddle{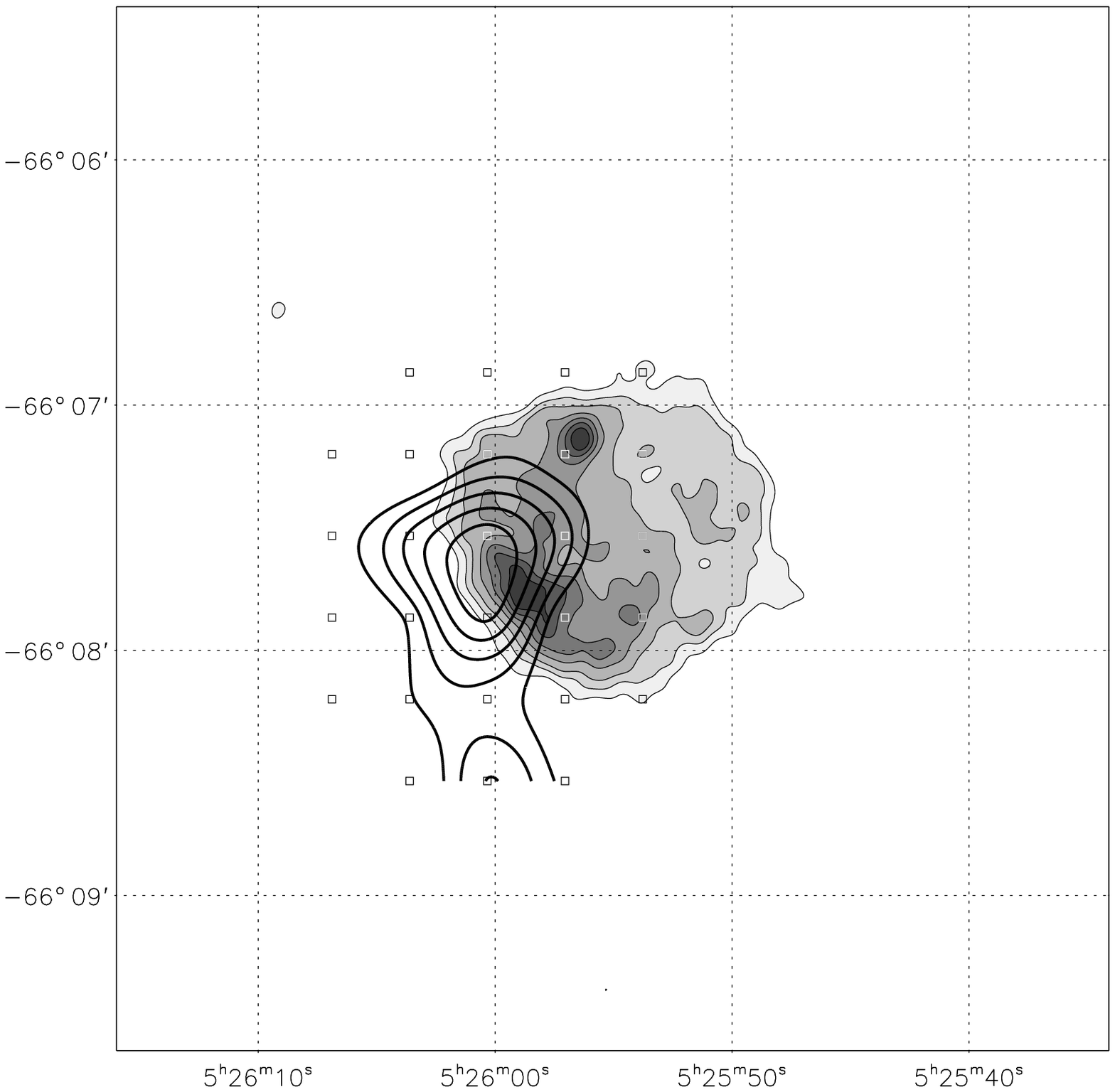}{4in}{0}{117}{117}{-340}{-225}
\end{figure}

\clearpage
\begin{figure}
\plotfiddle{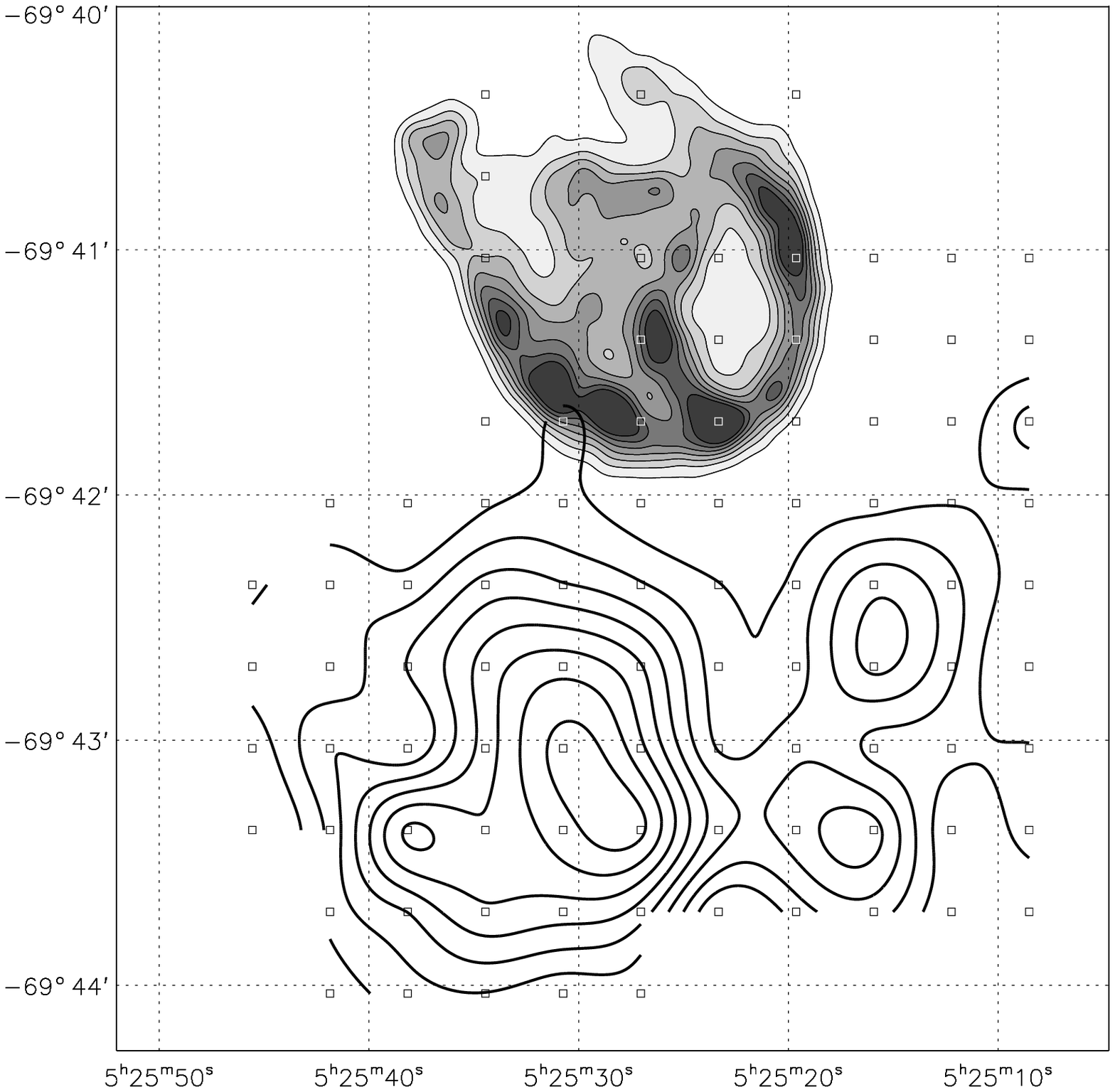}{4in}{0}{117}{117}{-340}{-225}
\end{figure}

\clearpage
\begin{figure}
\plotfiddle{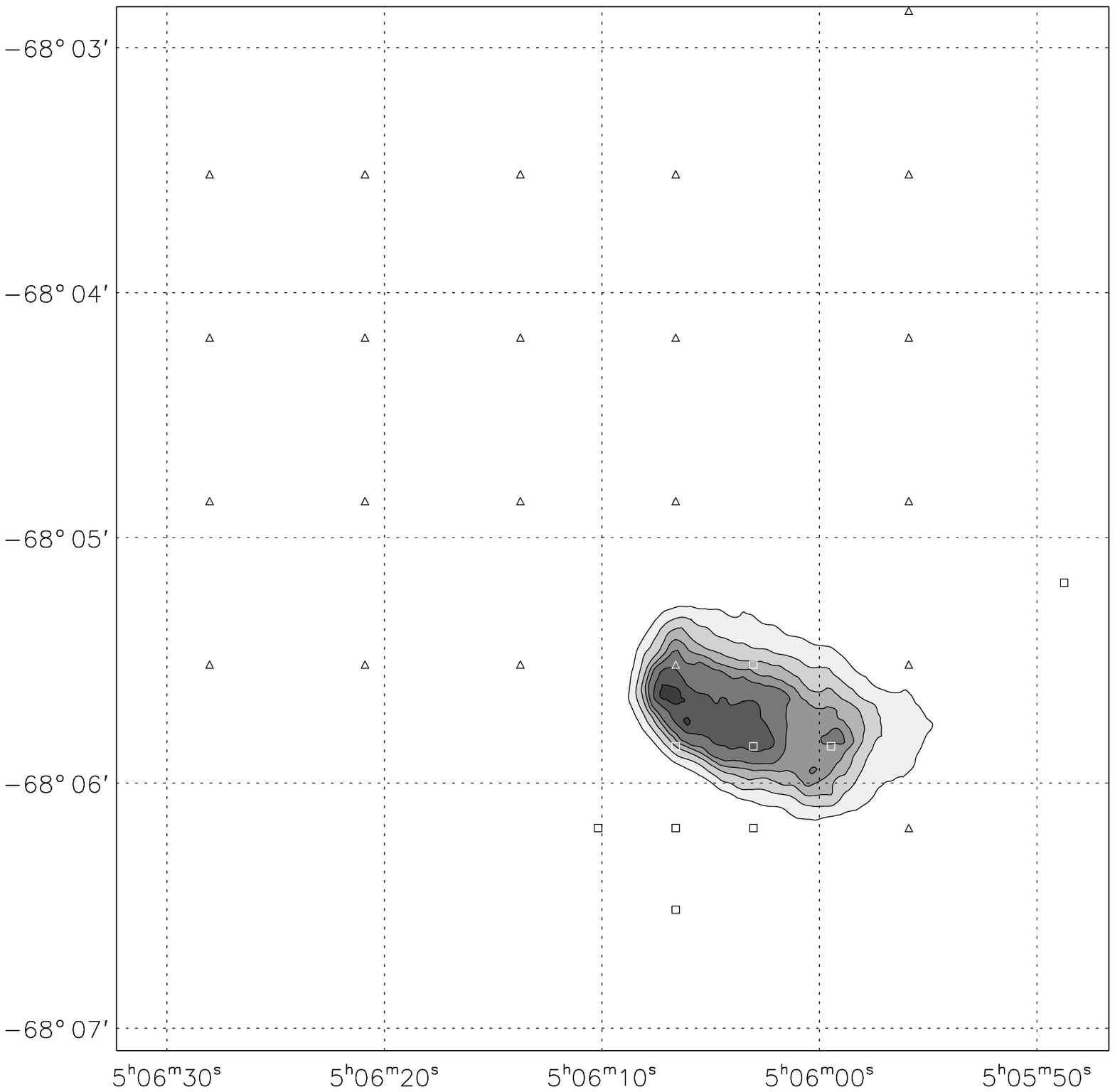}{4in}{0}{117}{117}{-340}{-225}
\end{figure}

\clearpage
\begin{figure}
\plotfiddle{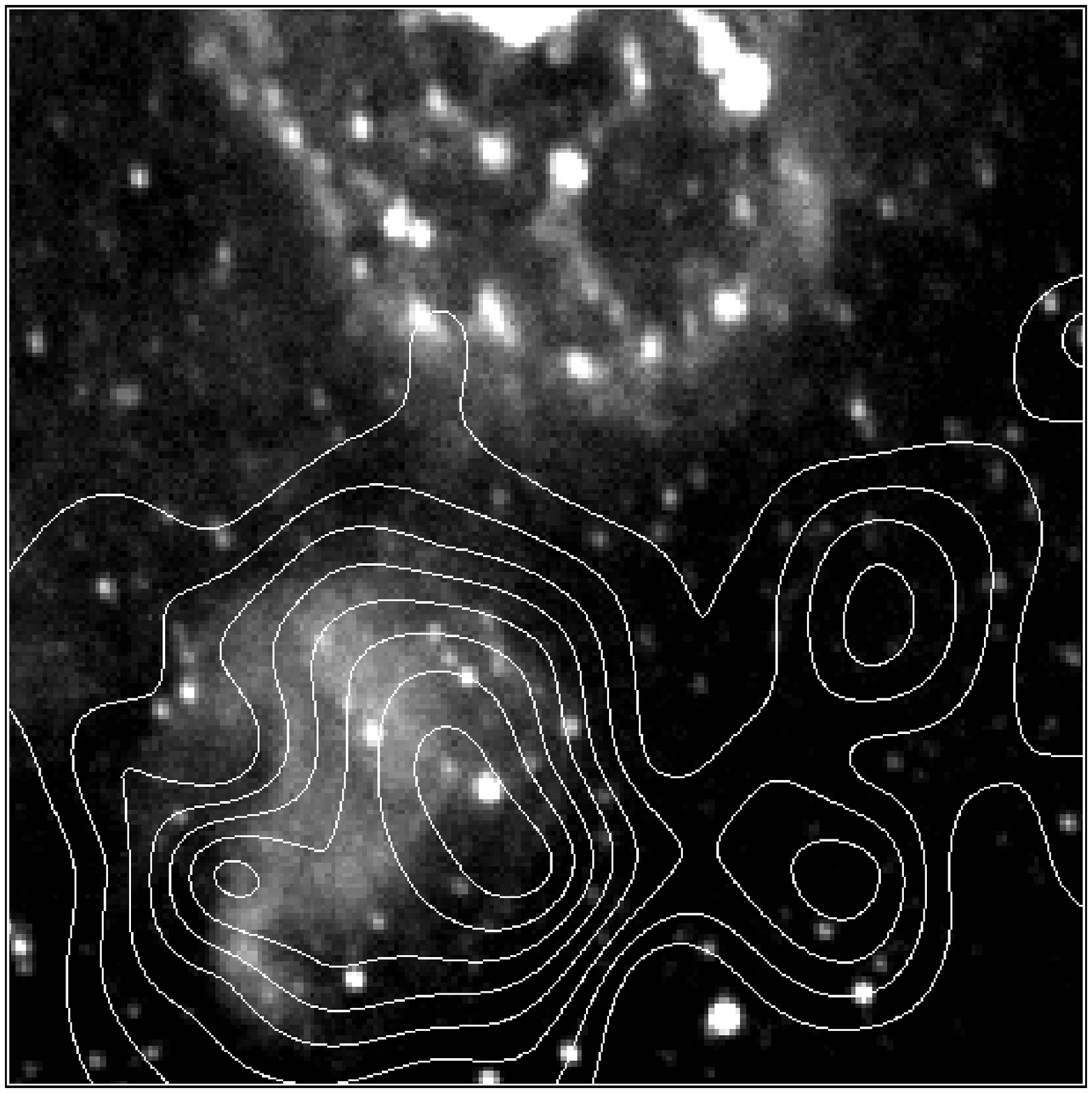}{4in}{0}{100}{100}{-300}{-270}
\end{figure} 

\end{document}